\documentclass[10pt,letterpaper]{article}
\usepackage[top=0.85in,left=2.75in,footskip=0.75in,marginparwidth=2in]{geometry}

\usepackage[utf8]{inputenc}

\usepackage{cite}
\usepackage{float}
\usepackage{ragged2e}

\usepackage{nameref,hyperref}
\usepackage{amsmath}%
\usepackage{amsfonts}%
\usepackage{amssymb}%
\usepackage[right]{lineno}

\usepackage{microtype}
\DisableLigatures[f]{encoding = *, family = * }

\raggedright
\usepackage{changepage}

\usepackage[aboveskip=1pt,labelfont=bf,labelsep=period,singlelinecheck=off]{caption}

\makeatletter
\renewcommand{\@biblabel}[1]{\quad#1.}
\makeatother

\usepackage{lastpage,fancyhdr,graphicx}
\usepackage{epstopdf}
\fancyheadoffset[L]{2.25in}
\fancyfootoffset[L]{2.25in}

\usepackage{color}

\definecolor{Gray}{gray}{.25}

\usepackage{graphicx}

\usepackage{sidecap}

\usepackage{wrapfig}
\usepackage[pscoord]{eso-pic}
\usepackage[fulladjust]{marginnote}
\reversemarginpar

\begin{document}
\vspace*{0.35in}

\begin{flushleft}
{\Large
\textbf\newline{\textbf{Bose-like few-fermion systems}}
}
\newline
\\
Yu-Lin Zhao\textsuperscript{1},
Chi-Chun Zhou\textsuperscript{2,1,\dag},
Wen-Du Li\textsuperscript{3,4,1,\dag\dag},
Wu-Sheng Dai\textsuperscript{1,*}
\\
\smallskip
\textsuperscript{1} Department of Physics, Tianjin University, Tianjin 300350, PR China
\\
\textsuperscript{2} School of Engineering, Dali University, Dali, Yunnan 671003, PR China
\\
 \textsuperscript{3}College of Physics and Materials Science, Tianjin Normal University, Tianjin 300387, PR China
\\
\textsuperscript{4}Theoretical Physics Division, Chern Institute of Mathematics, Nankai University, Tianjin, 300071, PR China

\bigskip
\textsuperscript{\dag} zhouchichun@dali.edu.cn\\
\textsuperscript{\dag\dag}liwendu@tju.edu.cn\\
* daiwusheng@tju.edu.cn

\end{flushleft}


\bigskip

\justifying \textbf{Dealing with a few-fermion system in the canonical ensemble, rather than in the grand canonical ensemble, shows that a few-fermion system with odd number fermions behaves differently from a few-fermion system with even number fermions. An even-number-fermion system behaves like a Bose system rather than a Fermi system.}
\newline


\textbf{Introduction} A system consisting of finite number particles, strictly speaking, should be
dealt with in the canonical ensemble. In quantum statistics, however, few-body
systems are always dealt with in the grand canonical ensemble. Considering
few-body systems in the grand canonical ensemble is indeed an
approximation.\ In the grand canonical ensemble, the accurate particle number
is approximated by the mean particle number. In this paper, by dealing with
few-fermion systems in the canonical ensemble rather than in the grand
canonical ensemble, we find the difference between odd-particle-number and
even-particle-number few-fermion systems: an even-particle-number few-fermion
system behaves like a Bose system.

The reason why approximately using the grand canonical ensemble instead of the
canonical ensemble to deal with few-particle systems is that in the canonical
ensemble the constraint on the particle number makes the calculation
complicated. As an expedient treatment, one turns to the grand canonical
ensemble which has no constraint on the particle number. Nevertheless, in the
grand canonical ensemble one cannot distinguish the difference between odd-
and even-particle-number Fermi systems, since in the grand canonical ensemble
there is no restriction on the particle number.\newline

The problem of few-body systems draws much interest. Few-body systems exhibit
obvious quantum effects at low temperatures. The technology today makes it
possible to implement and operate a cold few-body system
\cite{chu1985three,metcalf2007laser,onofrio2017physics,sowinski2019one}. Cold and ultracold
few-body systems are discussed
\cite{wang2013ultracold,d2018few,pollack2009universality}, and the cold and
ultracold few-fermion systems
\cite{yannouleas2016ultracold,zollner2006ultracold} are studied intensely.
Few-fermion systems in one-dimensional
\cite{wang2012absence,rammelmuller2015few}, the effective theory for trapped
few-fermion systems \cite{stetcu2007effective}, and strongly interacting
few-fermion systems \cite{thomas2010nearly,wu2011strongly} are discussed.
There are experiments on cold and ultracold few-fermion systems
\cite{fukuhara2007degenerate,desalvo2010degenerate,martiyanov2010observation,lu2012quantum,stas2004simultaneous,vassen2012cold}%
. The experiments in one and two dimensions
\cite{martiyanov2010observation,orel2011density,guan2013fermi,makhalov2014ground}
and the thermal quantities of cold and ultracold few-fermion systems
\cite{stewart2006potential,luo2009thermodynamic,luo2007measurement,kinast2005heat,horikoshi2010measurement}
are reported. There are studies on two-body correlations for few-body systems
\cite{pecak2019intercomponent} and Cooper-like Fermi-Fermi mixtures
\cite{sowinski2015slightly}. Few-fermion thermodynamics is also studied
\cite{armstrong2012virial,armstrong2013thermodynamics,armstrong2012quantum}.

\textbf{Ideal Fermi gases in the canonical ensemble.} In the canonical
ensemble, the canonical partition function of a $\nu$-dimensional ideal Fermi
gas with $N$ fermions is \cite{zhou2018canonical}%
\begin{equation}
Z_{FD}\left(  \beta,N\right)  =\frac{Z^{N}\left(  \beta\right)  }{N!}\det
M_{N} \label{Zbeta}%
\end{equation}
with%
\begin{equation}
M_{N}=\left(
\begin{array}
[c]{cccc}%
1 & \frac{1}{Z\left(  \beta\right)  } & \cdots & 0\\
\frac{1}{2^{\nu/2}} & 1 & \cdots & 0\\
\vdots & \vdots & \ddots & \frac{N-1}{Z\left(  \beta\right)  }\\
\frac{1}{N^{\nu/2}} & \frac{1}{\left(  N-1\right)  ^{\nu/2}} & \cdots & 1
\end{array}
\right)  ,
\end{equation}
where $Z\left(  \beta\right)  =\sum\nolimits_{s}e^{-\beta\epsilon_{s}}%
=g\frac{V}{\lambda^{\nu}}$ is the single-particle partition function of an
ideal classical gas, $V$ is the volume, $g$ is the number of internal degrees
of freedom, and $\lambda=h/\sqrt{2\pi mkT}$ is the thermal wave length.
Various thermodynamic quantities in the canonical ensemble can be obtained
from the partition function (\ref{Zbeta}), e.g., the specific heat%
\begin{equation}
\frac{C_{V}}{Nk}=\frac{\nu}{2}+\frac{1}{N}\left(  2T\frac{\partial}{\partial
T}\ln\det M_{N}+T^{2}\frac{\partial^{2}}{\partial T^{2}}\ln\det M_{N}\right)
. \label{CV}%
\end{equation}
The first term in the specific heat, $\frac{\nu}{2}$, is the specific heat of
the corresponding classical gas, and the rest part is the quantum effect
coming from the exchange interaction.\newline

\textbf{Ideal Fermi gases in the grand canonical ensemble.}\textit{ }As a
comparison, consider the above few-fermion system in the grand canonical
ensemble. In the grand canonical ensemble, in stead of the canonical partition
function (\ref{Zbeta}), one works with the grand partition function $\Xi$,
given by $\ln\Xi=\sum\nolimits_{s}\ln\left(  1+ze^{-\beta\epsilon_{s}}\right)
=g\frac{V}{\lambda^{\nu}}f_{\nu/2+1}\left(  z\right)  $ with $f_{\sigma
}\left(  z\right)  $ the Fermi-Dirac integral \cite{pathria2011statistical}.
The calculation of\ the grand partition function is simple, because there is
no limitation on the particle number. In doing so, one losses the information
of the particle number. For a finite-size system, however, one need to take
the particle number into account. For this purpose, one requires that the mean
particle number $\left\langle N\right\rangle $ equals the exact particle
number $N$. The restriction $\left\langle N\right\rangle =N$\ gives the
relation between the chemical potential and the temperature
\cite{dai2017explicit}. As a result, in the grand canonical ensemble, one
cannot distinguish the difference caused by the odevity of particle numbers.

In the grand canonical ensemble, the specific heat reads
\cite{pathria2011statistical}%
\begin{equation}
\frac{C_{V}}{Nk}=\frac{\nu}{2}\left(  \frac{\nu+2}{2}\frac{f_{\nu/2+1}\left(
z\right)  }{f_{\nu/2}\left(  z\right)  }-\frac{\nu}{2}\frac{f_{\nu/2}\left(
z\right)  }{f_{\nu/2-1}\left(  z\right)  }\right)  , \label{gFDCv}%
\end{equation}
where the fugacity is given by $\ln z\simeq\frac{\varepsilon_{F}}{kT}\left[
1-\frac{\pi^{2}}{12}\left(  \frac{kT}{\varepsilon_{F}}\right)  ^{2}\right]  $
with $\varepsilon_{F}=\frac{h^{2}}{2m}\left(  \frac{3n}{4\pi g}\right)
^{2/3}$ the Fermi energy.

\textbf{The lower limit on the temperature of a finite size system.}\textit{
}Few-particle systems must be of a finite size, and the wavelength of particles must be less than the system size. Therefore, for finite-size
systems, there exists a lowest temperature, depending on the system size.

Consider a particle in a container of size $L$. The one-half wavelength should
be less than $L$, i.e., $\lambda/2\lesssim L$. This gives a constraint on the
temperature, $T\gtrsim\frac{1}{4}\frac{h^{2}}{2\pi mk}\frac{1}{L^{2}}$, or,
equivalently, $T\gtrsim\frac{1}{4}\frac{h^{2}}{2\pi mk}n^{2/3}$, where
$n\sim1/L^{3}$ is the particle number density.\newline

\begin{wrapfigure}[15]{l}[0cm]{55mm}
\includegraphics[width=55mm]{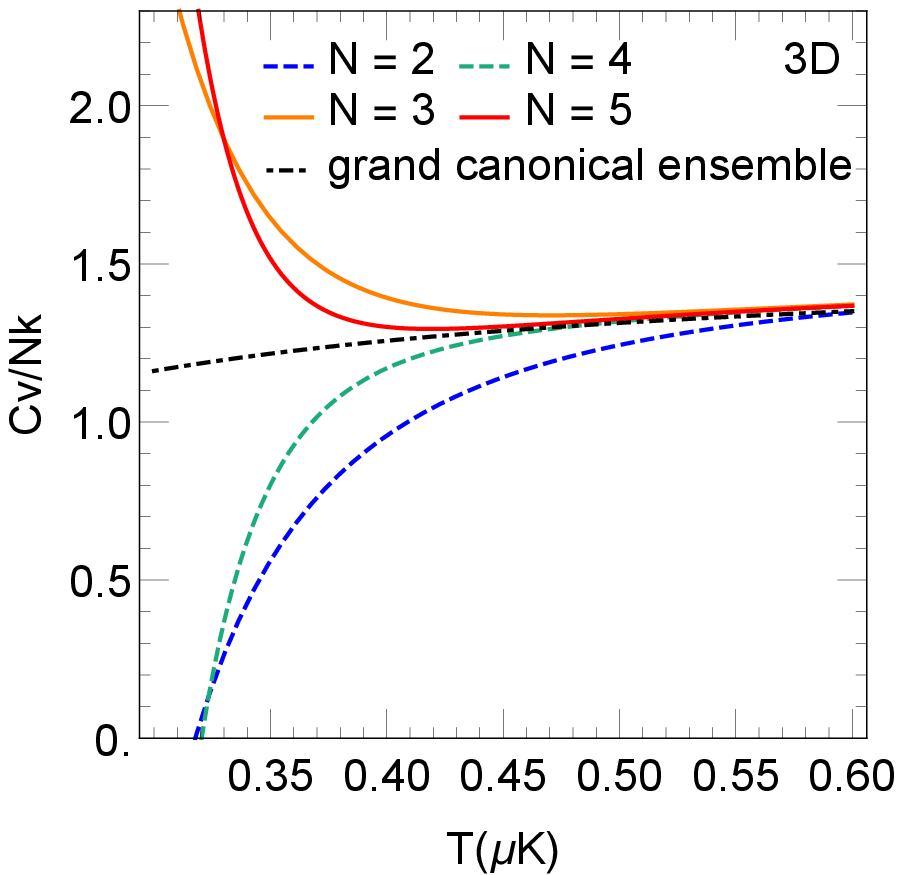}
\captionsetup{labelformat=empty} 
\caption{} 
\label{3D-3He-Cv} 
\end{wrapfigure}

\newpage
\marginpar{
\vspace{0.8cm} 
\color{Gray} 
\textbf{Figure \ref{3D-3He-Cv}. The heat capacity of a 3D Fermi gas.} 
In three dimensions, the odd-particle-number few-fermion system and the even-particle-number one behave differently.
}

\textbf{The }$^{3}$\textbf{He gas: an example.}\textit{ }The $^{3}$He atom, a
fermion with spin\ $1/2$ and mass $5.019\times10^{-27}kg$, plays an important
role in cold atom experiments \cite{cohen2018mecanique}. There are many
experiments on the the cold and ultracold $^{3}$He gases
\cite{stas2004simultaneous,vassen2012cold}. Consider a\ $^{3}$He gas with the
density $n\simeq1\times10^{18}m^{-3}$
\cite{leggett2006quantum,mcnamara2006degenerate,notermans2016comparison}. The
lower limit on the temperature, by the above estimation, is about $0.25\mu K$.

\begin{wrapfigure}[15]{l}[0cm]{55mm}
\includegraphics[width=55mm]{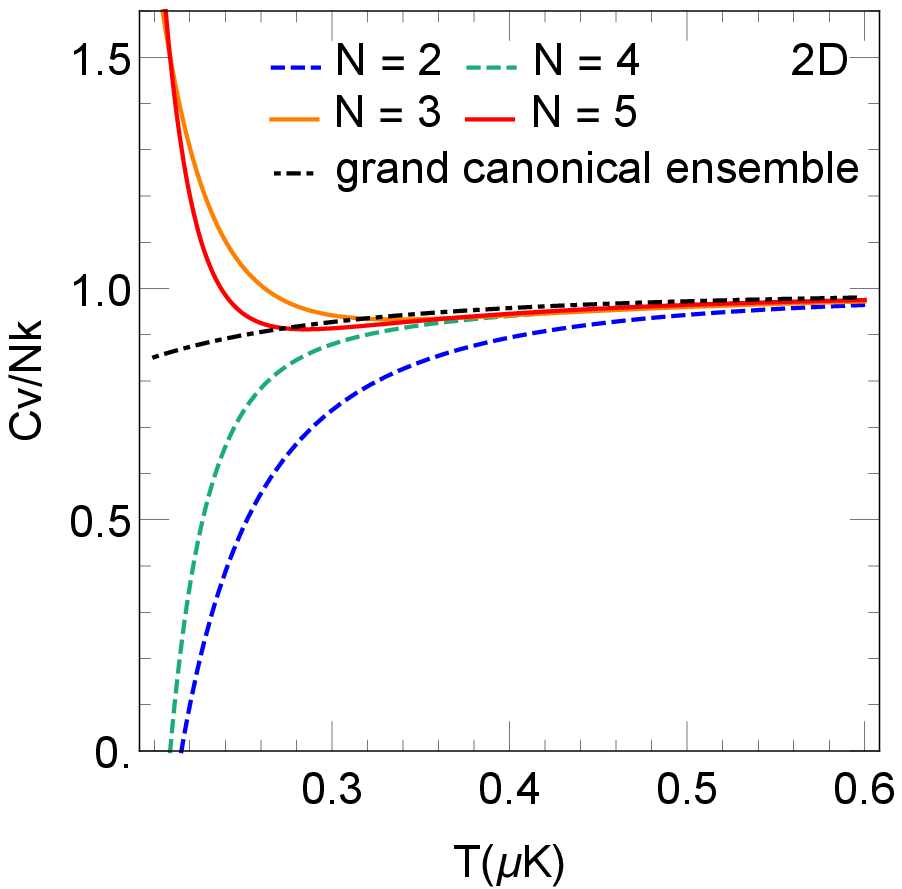}
\captionsetup{labelformat=empty} 
\caption{} 
\label{2D-3He-Cv} 
\end{wrapfigure}

\marginpar{
\vspace{5.5cm} 
\color{Gray} 
\textbf{Figure \ref{2D-3He-Cv}. The heat capacity of a 2D Fermi gas.} 
In two dimensions, the odd-particle-number few-fermion system and the even-particle-number one behave differently.
}

Consider the specific heat of a $^{3}$He gas with the particle number $N=2$,
$3$, $4$, and $5$ in the canonical ensemble for roughly illustrating the
problem. As a comparison, we also provide the specific heat of the $^{3}$He
gas in the grand canonical ensemble.

\begin{wrapfigure}[15]{l}[0cm]{55mm}
\includegraphics[width=55mm]{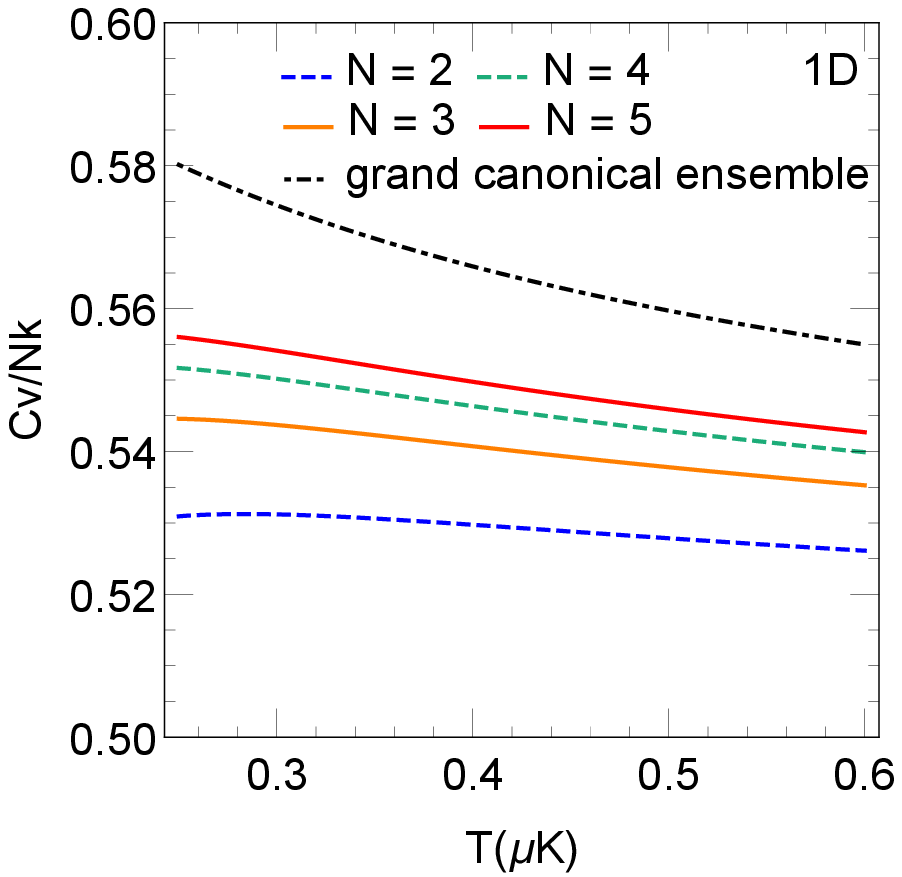}
\captionsetup{labelformat=empty} 
\caption{} 
\label{1D-3He-Cv} 
\end{wrapfigure}

Figures (\ref{3D-3He-Cv}) to (\ref{1D-3He-Cv}) show that the few-fermion gas
of odd-number atoms and the few-fermion gas of even-number atoms behave
differently at low temperatures: as the temperature decreases, the heat
capacity of the odd-number-fermion system increases while the heat capacity of
the even-number-fermion system decreases.

Nevertheless, in the grand canonical ensemble, the behaviors of odd- and
even-number-fermion systems are the same. The reason, as mentioned above, is
that in the grand canonical ensemble the information of the odevity of the
particle number is lost.

\marginpar{
\vspace{5.6cm} 
\color{Gray} 
\textbf{Figure \ref{1D-3He-Cv}. The heat capacity of a 1D Fermi gas.} 
In one dimension, the odd-particle-number few-fermion system and the even-particle-number one behave similarly.
}

Finally compare few-fermion systems with few-boson systems.\ In order to
facilitate comparison, the fictitious Bose system is chosen as\ a Bose system
with the same parameters as the $^{3}$He gas. The result shows that the
specific heat of few-boson systems behaves like the specific heat of
even-fermion systems, see Figure (4). It should be emphasized that a
three-dimensional ideal Bose gas may perform the BEC phase transition. This is
the reason why there is a sharp peak in the specific heat of the
three-dimensional Bose gas calculated in the grand canonical ensemble. In
contrast, a finite-particle-number system cannot perform phase transitions
\cite{pathria2011statistical}, so the specific heat calculated in canonical
ensemble has no sharp peak but only has a maximum value.

\textbf{A word on spatial dimensions.} Figures (\ref{3D-3He-Cv}) to
(\ref{1D-3He-Cv}) show that the specific heat in
one, two, and three
dimensions behave differently. This is because exchange interactions in
different dimensional space are different. The exchange interaction in higher
dimensions is more obvious than that in lower dimensions. The difference
between the quantum gas and the classical gas is caused by the exchange
interaction. Therefore, the difference between odd- and even-particle-number
systems in higher dimensions is more obvious than the difference in lower dimensions.

\vspace{1cm} 
\begin{adjustwidth}{-3.5in}{0in}
\begin{flushright}
\includegraphics[width=156mm]{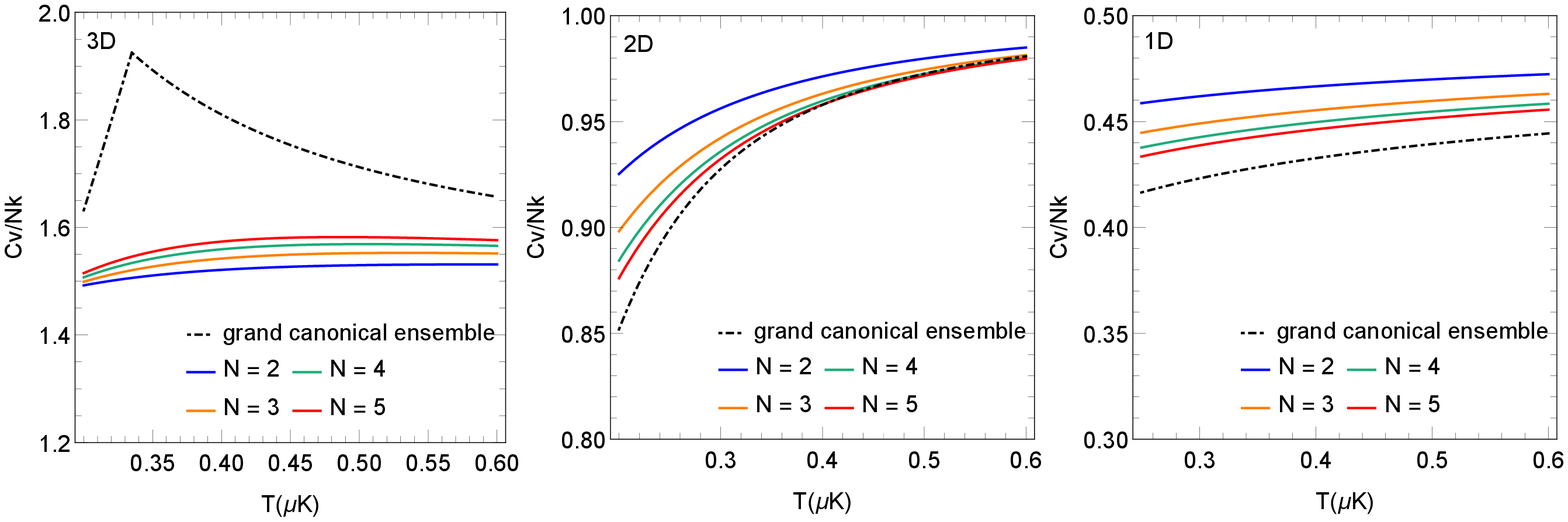}
\end{flushright}
\justify
\color{Gray}
\textbf {\qquad \qquad \qquad \qquad \qquad  \qquad \qquad \qquad \qquad Figure 4. The heat capacity of a 3D Bose gas.}
\end{adjustwidth}



\textbf{Summary}
At low temperatures, odd-particle-number few-fermion systems
behave differently from even-particle-number few-fermion systems. The specific
heat of an even-particle-number few-fermion system behaves like a Bose system
rather than a Fermi system. This inspire us to simulate a few-boson system by
a few-fermion system. One can switch a Bose-like system to a Fermi system and
\textit{vice versa} by changing the number of fermions.



\section*{Acknowledgments}
We are very indebted to Dr G. Zeitrauman for his encouragement. This work is supported in part
by NSF of China under Grant No. 11575125 and No. 11675119.

\nolinenumbers


\bibliographystyle{unsrt}

\end{document}